# Eguard: Defending LLM Embeddings Against Inversion Attacks via Text Mutual Information Optimization

**Tiantian Liu[1,2,3], Hongwei Yao[1,2], Feng Lin[1,2], Tong Wu[1,2], Zhan Qin[1,2], Kui Ren[1,2]**

[1]State Key Laboratory of Blockchain and Data Security, Zhejiang University
[2]Hangzhou High-Tech Zone (Binjiang) Institute of Blockchain and Data Security
[3]School of Informatics, Xiamen University
{tiantian, yhongwei, flin, cocotwu, qinzhan, kuiren}@zju.edu.cn

## Abstract

While text embeddings enable efficient semantic processing in LLMs, they remain vulnerable to inversion attacks that reconstruct sensitive original text. However, current defense methods typically treat text embeddings from the feature level independently, ignoring the exploitation of the mutual relation among the embedding construction pipeline. To address this limitation, we propose `Eguard`, a framework that effectively disrupts chains of relationships between the original semantic space and defended functional space. Our improvements manifest at two levels, i.e., the global-level and local-level mutual information. At the global level, we propose to minimize the statistical dependency between protected embeddings and their original inputs, effectively decoupling sensitive content from the semantic space accessible to adversaries. At the local level, we apply keyword-antonym contrastive learning to enforce semantic discriminability within the space of downstream utility. This synergy of global privacy control and local semantic alignment allows `Eguard` to achieve a superior privacy-utility trade-off than traditional defenses. Our approach significantly reduces privacy risks, protecting over 95% of tokens from inversion while maintaining high performance across downstream tasks consistent with original embeddings.

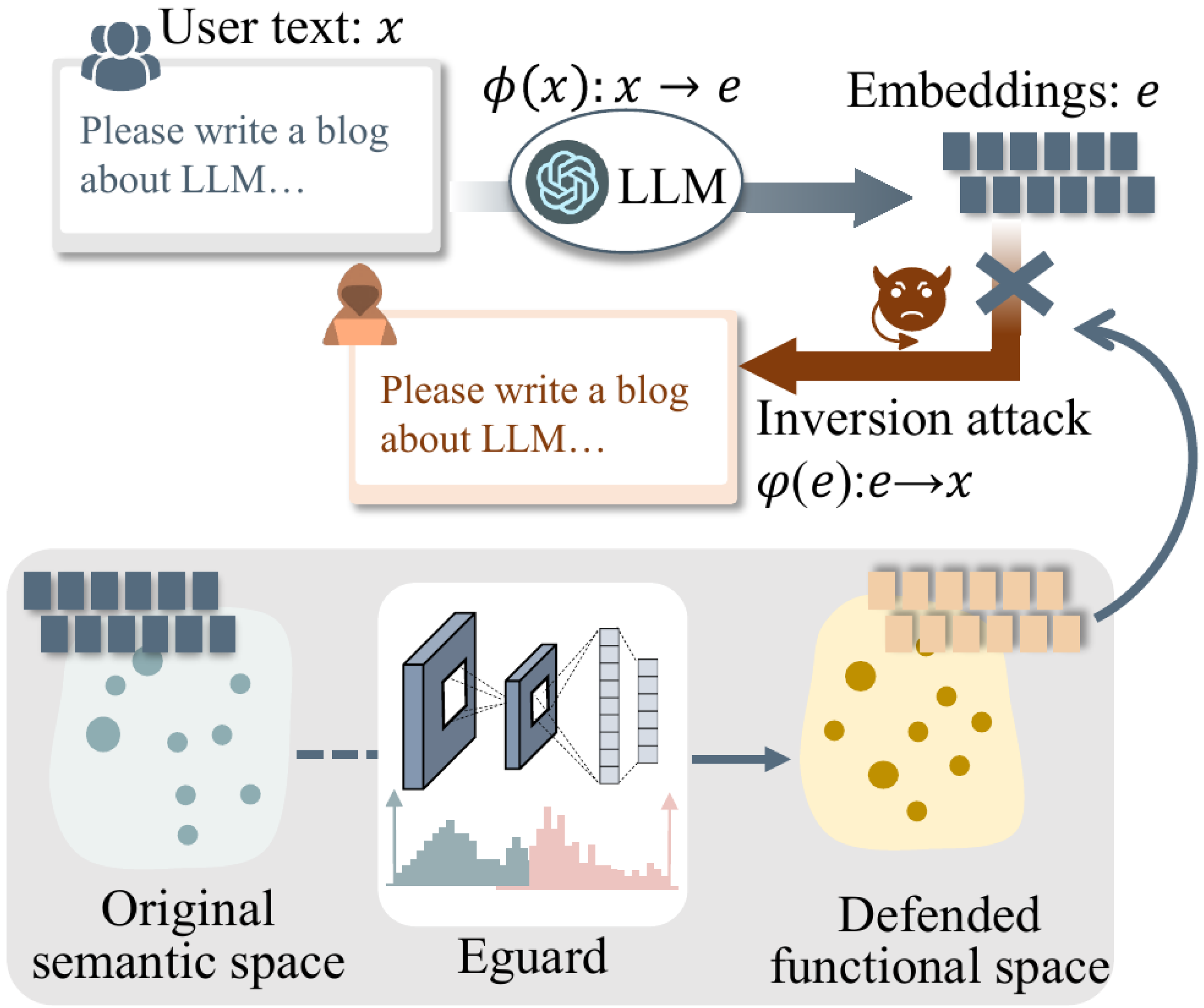

Figure 1: Overview of our defense against inversion attacks on embeddings.

## Introduction

Large language models (LLMs) like ChatGPT, Claude (Anthropic 2024), and ChatGLM rely critically on embedding vector databases as their long-term memory(Topsakal and Akinci 2023; Ozdemir 2023), enabling retrieval-augmented generation (RAG) (Gan et al. 2024; Zhao et al. 2024) to enhance factual accuracy and overcome context window limitations. These dense vector representations power essential capabilities including semantic search, clustering, and context-aware generation, as demonstrated by widespread adoption in APIs like OpenAI's embeddings service. However, the richness of semantic information that makes embeddings valuable also creates privacy vulnerabilities.

Recent studies have revealed that embeddings are vulnerable to *inversion attacks* (Lin et al. 2024; Li, Xu, and Song 2023; Kale et al. 2025), where adversaries exploit embedding vectors to reconstruct sensitive input texts. Such attacks pose severe risks in scenarios involving proprietary data or personal information. Researchers have proposed several defense mechanisms to mitigate these risks. These defenses can generally be categorized into three main approaches: noise superposition (Wen, Yiu, and Hui 2021; Liu et al. 2024), perturbation-based methods (Jin et al. 2025), and differential privacy (DP)-based defenses (Wang et al. 2024; Li et al. 2024). Despite these efforts, existing defense methods face significant limitations that hinder their practical effectiveness against embedding inversion attacks. Defending against such attacks presents unique challenges: First, embedding vectors inherently encapsulate sensitive input features, making it difficult to disentangle and protect private information without severely disrupting their semantic integrity. Second, modifying embedding layers often compromises the accuracy of LLMs on downstream tasks. Given the versatility of LLMs across applications such as sentiment analysis, natural language inference, and text summarization, developing robust defense mechanisms that simultaneously ensure privacy protection while preserving task-specific utility remains a critical research problem.

The embedding-to-text reconstruction process can be viewed as a Markov chain, and we mitigate inversion attacks by introducing an additional stage that disrupts the correlation between original text and reconstructed outputs. In this paper, we propose `Eguard` (**Embedding Guard**), a novel defense mechanism that leverages mutual information optimization to shield embeddings from inversion attacks while preserving their effectiveness for natural language processing (NLP) tasks. The optimization process is driven by text mutual information, ensuring that the transformed embeddings retain task-relevant properties while minimizing their susceptibility to inversion. Specifically, `Eguard` introduces a projection network that acts as an intermediate stage between the original embedding and the reconstructed text, increasing the uncertainty of inversion attacks by leveraging information entropy principles. A fundamental challenge in this defense lies in decoupling sensitive information from projected embeddings, thereby significantly reducing the adversary's ability to recover original inputs. `Eguard` addresses this by leveraging global mutual information as a quantitative measure of relationships on the chain, systematically minimizing its approximation to sever the linkage between sensitive data and embeddings, thereby preventing information leakage. Additionally, protecting embeddings alone is not enough; their functional utility for downstream tasks must also be maintained. To address this, `Eguard` introduces a local mutual information optimization mechanism that enforces semantic discriminability through keyword-antonym contrastive learning. By maximizing the contrastive separation between keywords and their antonyms within a conditional distribution space, the model ensures that the projected embeddings preserve essential semantic distinctions while remaining anchored in the feature space necessary for robust task performance.

Our key contributions are summarized as follows:

- We propose `Eguard`, a transformer-based projection network that mitigates inversion attacks by projecting embeddings into a secure latent space while preserving downstream performance.
- We develop a mutual information-driven framework that simultaneously minimizes global information leakage and maintains local semantic discriminability, ensuring robust yet useful embeddings.
- We conduct comprehensive evaluations on seven embedding models and two ChatGPT embeddings under various attacks, demonstrating that `Eguard` blocks over 95% of token inversion, maintains more than 98% task consistency, and shows strong robustness to perturbations, unseen datasets, and adaptive attacks.

## Embedding Models and Inversion Threats

An embedding model transforms an input text $x = [w_1, \ldots, w_l]$ from vocabulary $\mathcal{V}$ into a compact vector $e \in \mathbb{R}^d$, capturing semantic and syntactic features. Each token is mapped via $\boldsymbol{v}_i = \mathbf{W}(w_i)$ and processed by a Transformer encoder (e.g., BERT (Devlin et al. 2018), T5 (Raffel et al. 2020), LLaMA (Touvron et al. 2023)) to produce hidden states $\boldsymbol{h} = [\boldsymbol{h}_1, \ldots, \boldsymbol{h}_l]$. A pooling operation (mean, max, or using the hidden state of a special token `[CLS]`) yields the final embedding $e = \text{Pooling}(\boldsymbol{h})$. These embeddings, serving as semantic representations, support applications in retrieval, classification, generation, and recommendation systems.

**Adversary and Threat Model:** Embedding inversion attacks attempt to reconstruct $x$ from its embedding $\phi(x)$ by building an inverse model $\varphi(e) : e \to x$. We assume a strong adversary capable of acquiring embeddings (e.g., via database breaches) and equipped with an auxiliary dataset $\mathcal{D}_{aux}$ matching the distribution of target texts. An adaptive adversary may also exploit knowledge of defensive mechanisms to refine attack strategies. Given $\phi(x)$, the attacker seeks a candidate text $\hat{x}$ minimizing the distance to the original embedding:

$$\hat{x} = \arg\min_{\hat{x} \in \mathcal{V}} \|\phi(\hat{x}) - \phi(x)\|_2. \tag{1}$$

Brute-force search is computationally infeasible due to combinatorial explosion. To overcome this, attackers employ a decoder-based transformer $\varphi$ (e.g., GPT-2) trained to maximize the likelihood:

$$\arg\max_{\varphi} \mathbb{E}_{x \sim \mathcal{D}_{aux}}[p(x|\varphi(\phi(x)))]. \tag{2}$$

The decoder generates tokens sequentially, modeling $p(x) = \prod_{i=1}^{n} p(t_i|\phi(x), t_{<i})$, with self-attention capturing long-range dependencies. A linear projection aligns embedding dimensions with the decoder input, enabling accurate text reconstruction from embeddings.

## `Eguard` Defense

In this section, we present `Eguard`, a novel defense mechanism based on mutual information optimization, designed to mitigate embedding inversion attacks while preserving the functionality of embeddings. `Eguard` achieves this by projecting embedding vectors from their original space into a secured space, guided by two key objectives: global mutual information and local mutual information. The reduction of global mutual information increases the uncertainty of reconstructing original texts from inversion attacks, thereby detaching sensitive information and enhancing privacy. Simultaneously, the local mutual information between keywords and their corresponding antonyms ensures the preservation of discriminative semantic features, maintaining the effectiveness of embeddings for downstream tasks in large language models.

### Global Mutual Information: Sensitive Feature Detachment

To effectively mitigate those inversion attacks, we propose an information-theoretic defense mechanism that reduces the dependency between original textual content and projected embeddings.

By analyzing the attack pipeline, we model the embedding generation and attacking process as a Markov chain: $x \to e \to \hat{x}$, where $x$ represents the original text, $e$ is its corresponding embedding, and $\hat{x}$ is the reconstructed text obtained through the attack. To enhance security, we introduce a projection network, which transforms the original

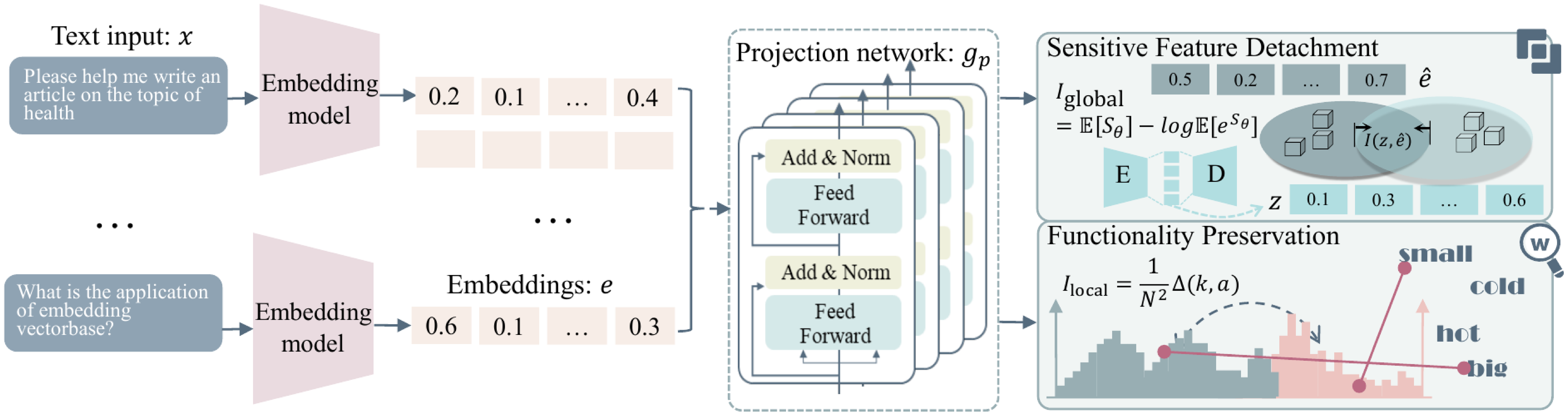

Figure 2: Overview of defense approach. `Eguard` contains a sensitive feature detachment module and a functionality preservation module.

embedding $e$ into a secured representation $\hat{e}$. Consequently, the Markov chain is extended as: $x \rightarrow e \rightarrow \hat{e} \rightarrow \hat{x}$. From an information-theoretic standpoint, the objective is to minimize the mutual information $I(x, \hat{e})$ between the original text $x$ and the projected embedding $\hat{e}$, thereby reducing the amount of recoverable information:

The mutual information is formally defined through Shannon entropy relationships:

$$I(x, \hat{e}) = H(x) - H(x|\hat{e}), \tag{3}$$

where $H(\cdot)$ denotes Shannon entropy, and $H(x|\hat{e})$ represents the conditional entropy of $x$ given $\hat{e}$. Prior research has established that mutual information can be approximated using the Kullback-Leibler (KL) divergence:

$$\begin{aligned} I(x, \hat{e}) &= \iint p(\hat{e}|x)p(x) \log \frac{p(x|\hat{e})}{p(\hat{e})} dx d\hat{e} \\ &= KL(p(\hat{e}|x)p(x) \| p(\hat{e})p(x)). \end{aligned} \tag{4}$$

Since $x$ represents discrete text data, it is not directly possible to calculate the mutual information between $x$ and a continuous vector $\hat{e}$. To address this, we introduce a pre-trained autoencoder that is first trained on a large and unlabeled text corpus. The latent feature vector $z$ produced by the autoencoder replaces $x$ to estimate the mutual information $I(z, \hat{e})$. According to information theory, the transformation between $z$ and $x$ is lossless only if the autoencoder can perfectly reconstruct the input $x$ from $z$. This implies that the conditional entropy $H(x|z)$ equals zero, ensuring that $z$ can effectively substitute $x$ in the calculation of mutual information.

Inspired by the work of Belghazi et al., we introduce a statistical network $S$ parameterized by $\theta$ to estimate the joint probability density function of $\hat{e}$ and $z$. This enables us to derive an ideal approximation of $I(z, \hat{e})$, which can be formulated as:

$$\begin{aligned} I_{\text{global}}(z, \hat{e}) = &\sup_\theta \iint S_\theta(z, \hat{e}) p(z, \hat{e}) dz d\hat{e} \\ &- \log \left( \iint e^{S_\theta(z, \hat{e})} dp(z) dp(\hat{e}) \right). \end{aligned} \tag{5}$$

Eq. 5 provides a rigorous estimation of global mutual information, and its computational implementation is detailed in Algorithm 1. The training objective incorporating this global mutual information is defined as:

$$\mathcal{L} = \ell_{\text{task}} + \alpha I_{\text{global}}(z, \hat{e}), \tag{6}$$

where $\ell_{\text{task}}$ represents the loss function associated with the downstream task, and $\alpha$ is a hyperparameter that controls the influence of global mutual information regularization.

**Lemma 1** *Minimizing the mutual information $I(x, \hat{e})$ between the original text $x$ and the secured embedding $\hat{e}$ effectively protects against inversion attacks by reducing the recoverable information about $x$ from the attacker's reconstructed text $\hat{x}$.*

**Proof of Lemma 1:** see Appendix A in supplementary materials for proof.

**Lemma 2.** *For any $\epsilon > 0$, there exists a neural network $S_\theta$ such that:*

$$|I(z, \hat{e}) - I_{\text{global}}(z, \hat{e})| \leq \epsilon, \quad \textit{almost surely.} \tag{7}$$

*This lemma demonstrates that the global mutual information estimation $I_{\text{global}}(z, \hat{e})$ converges to the neural-based mutual information measure $I(z, \hat{e})$ as the number of training samples tends to infinity, ensuring the approximation error is bounded by $\epsilon$ with probability one.*(Belghazi et al. 2018)

**Proof of Lemma 2:** see Appendix B in supplementary materials for proof.

## Local Mutual Information: Functionality Preservation

While global mutual information ensures robustness against inversion attacks, it introduces a crucial challenge, that is, preserving the effectiveness of embeddings for downstream tasks. To address this, we propose local mutual information as a mechanism for functionality preservation. The methodology begins by decomposing the input text into a set of keywords and retrieving their corresponding antonyms, by leveraging NLTK toolkits in Python for automated extraction. The keywords $k$ capture essential semantics and the antonyms $a$ represent opposite context. These pairs $(k, a)$ form the basis for contrastive learning, where the objective is to minimize the mutual information $I(k, a)$ between keywords and their antonyms. Formally, the upper bound on

Algorithm 1: Global and Local Mutual Information Optimization

**Require:** Original embeddings $e$, textual input $x$, pretrained autoencoder, projection network $g_p$, embedding encoder, statistical network $S$, and variational network $V$.
**Ensure:** Optimized projection network $g_p$ and refined embeddings $\hat{e}$.

```
Initialize:
Encode input text into latent representations:
z = Autoencoder(x)
Extract keywords and antonyms:
k, a = Encoder(NLTK(x))
for each training iteration do
    k ← g_p(k), a ← g_p(a);
    Estimate global mutual information:
    I_global = (1/N) Σ_{i=1}^N S(g_p(e_i), z_i) − log((1/N) Σ_{i=1}^N e^{S(g_p(e_i), z_i)});
    Estimate local mutual information:
    I_local = (1/N^2) Σ_{i=1}^N Σ_{j=1}^N [log V(a_i, k_i) − log V(a_j, k_i)]
    Update statistical network parameters:
    θ_S ← θ_S + η ∇_{θ_S} I_global
    θ_V ← θ_V + η ∇_{θ_V} (1/N) Σ_{i=1}^N log V(a_i, k_i);
    Optimize the projection network:
    g_p ← g_p − η ∇_{g_p} (ℓ_task + α I_global + β I_local)
end for
```

their mutual information is given by:

$$I_{\text{local}}(k, a) = \iint p(k, a) \log p(k|a), dk, da - \int p(k) \left[ \int p(a) \log p(a|k), da \right] dk. \quad (8)$$

**Lemma 3.** *For any keyword-antonym distribution, $I(k, a) \leq I_{\text{local}}(k, a)$, with equality if and only if k and a are independent.* (Cheng et al. 2020)

**Proof of Lemma 3:** see Appendix C in supplementary materials for proof.

The practical estimation of $I_{\text{local}}(k, a)$ involves training a variational network $V_\theta(k, a)$ to approximate $\log p(k|a)$, implemented via contrastive learning. For $N$ samples $\{(k_i, a_i)\}$, the objective becomes:

$$I_{\text{local}} = -\beta \cdot \mathbb{E} \left[ \log \frac{e^{V_\phi(k_i, a_i)}}{\int e^{V_\theta(k_i, a)} p(a)\, da} \right] \approx -\beta \cdot \frac{1}{N^2} \sum_{i,j} \left[ \log V_\theta(k_i, a_i) - \log V_\theta(k_i, a_j) \right], \quad (9)$$

where $p(a_i|k_i)$ represents the conditional log-likelihood of a positive sample pair $(k_i, a_i)$ and $p(a_j|k_i)$ represents the conditional log-likelihood of a negative sample pair $(k_i, a_j)$. The difference between the two terms reflects a contrastive probability log-ratio, ensuring that the model effectively distinguishes between meaningful keyword-antonym pairs. The local mutual information guarantees that the embeddings retain discriminative semantic features.

The unified training objective integrates global privacy protection and local utility preservation:

$$\mathcal{L} = \ell_{\text{task}}(y, f(g_p(e))) + \alpha I_{\text{global}}(z, \hat{e}) + \beta I_{\text{local}}(k, a). \quad (10)$$

where task-specific heads $f(\cdot)$ adapt to diverse applications: cross-entropy loss for classification, multiple negatives ranking loss (MNRL) for retrieval, and sequence cross-entropy for text generation. This formulation theoretically guarantees that sanitized embeddings $\hat{e}$ maintain sufficient statistics for downstream tasks while obfuscating extraneous information vulnerable to inversion attacks. The proposed Algorithm 1 systematically enhances embedding security while preserving their functional utility. The algorithm begins by encoding the input text into latent representations and extracting keywords and their antonyms. During each training iteration, the global mutual information is estimated using the statistical network $S$, while the local mutual information is computed via the variational network $V$. The parameters of $S$ and $V$ are updated to refine their estimates, and the projection network is optimized using gradient-based updates to balance privacy protection and task performance. This iterative process ensures that the embeddings are both secure and effective for downstream tasks. The projection network $g_p$ is implemented using a 24-layer RoBERTa encoder, which serves as an intermediate transformation module between raw embeddings and the secured space.

# Evaluation

## Experimental Setup

**Datasets and tasks.** We evaluate on four representative text understanding tasks: sentiment analysis SST, natural language inference (NLI), question retrieval (QR), and text summarization (TS). Each dataset is encoded by the embedding models for classification, retrieval, or summarization. Details of the datasets are provided in Appendix E of supplementary materials.

**Embedding model and Attack model.** We use five models: T5, RoBERTa, MPNet, LLaMA, and Gemma, all with frozen parameters. T5 and MPNet produce 768-dimensional embeddings, while RoBERTa outputs 1024-dimensional vectors. Model details are in Appendix D. Some additional results for Table 1 and Table 2 are included in the Appendix due to space limits. We adopt GPT-2 as the inversion decoder, with a maximum sequence length of 128. The decoder is trained on in-domain data using Adamax with a learning rate of 2e−2 and a batch size of 16.

**Metrics.** Defense performance is measured using F1, Recall, and BLEU, while task utility is assessed via Accuracy for SST, NLI, and QR, and ROUGE for TS. Lower F1, Recall, and BLEU indicate stronger protection, while higher task accuracy reflects better preservation of utility.

## Overall Performance

**Defense against embedding inversion attacks.** We evaluate the effectiveness of our defense across five embedding models, including T5, RoBERTa, MPNet, LLaMA and Gemma, and compare it with adversarial training methods FreeLB (Zhu et al. 2019) and FGSM (Kim 2020), as well as differential privacy approaches DPSanitization (Tong et al. 2024) and DPForward (Du et al. 2023; Zhang et al. 2025). The baseline row in Table 6, labeled W attack, highlights the vulnerability of embeddings where inversion success rates exceed 95%. FreeLB and FGSM provide only partial protection, with F1 and Recall dropping to 15–50%, due to their perturbations failing to fundamentally reshape the semantic space exploited by attackers. In contrast, differential privacy achieves stronger robustness with F1 and Recall between 9% and 29% but often reduces the quality of embeddings for downstream tasks. Our defense significantly outperforms these methods, lowering inversion success to roughly 4% and protecting over 95 percent of tokens, demonstrating a more substantial disruption of attacker inference.

| Model | Method | SST2 | | | NLI | | | QR | | | TS | | |
|---|---|---|---|---|---|---|---|---|---|---|---|---|---|
| | | F1(%) | Recall(%) | BLEU | F1(%) | Recall(%) | BLEU | F1(%) | Recall(%) | BLEU | F1(%) | Recall(%) | BLEU |
| T5 | W/ Attack | 93.9 | 93.3 | 0.836 | 96.5 | 95.0 | 0.789 | 98.2 | 97.9 | 0.976 | 95.2 | 94.7 | 0.901 |
| | FGSM | 14.2 | 16.1 | 0.092 | 25.9 | 19.4 | 0.121 | 36.3 | 34.4 | 0.230 | 39.3 | 38.6 | 0.245 |
| | FreeLB | 39.8 | 39.4 | 0.433 | 42.2 | 44.9 | 0.268 | 46.4 | 43.6 | 0.278 | 49.0 | 48.8 | 0.312 |
| | DPforward | 9.35 | 11.9 | 0.054 | 23.2 | 17.4 | 0.139 | 21.7 | 16.4 | 0.089 | 21.5 | 20.3 | 0.100 |
| | Sanitization | 6.75 | 11.0 | 0.030 | 23.2 | 16.3 | 0.095 | 23.5 | 21.7 | 0.103 | 22.6 | 22.1 | 0.092 |
| | **Ours** | **4.75** | **4.40** | **0.019** | **5.35** | **4.47** | **0.034** | **3.57** | **4.14** | **0.014** | **3.56** | **4.44** | **0.011** |
| LLaMA2 | W/ Attack | 93.9 | 93.1 | 0.831 | 83.3 | 81.1 | 0.948 | 98.5 | 98.1 | 0.985 | 96.9 | 95.9 | 0.914 |
| | FGSM | 14.2 | 16.1 | 0.092 | 43.2 | 34.5 | 0.352 | 37.9 | 36.6 | 0.237 | 38.8 | 37.3 | 0.218 |
| | FreeLB | 44.3 | 43.6 | 0.446 | 41.1 | 34.2 | 0.351 | 50.6 | 49.6 | 0.289 | 47.1 | 46.9 | 0.283 |
| | DPforward | 12.2 | 13.0 | 0.058 | 25.4 | 21.9 | 0.115 | 25.7 | 24.3 | 0.121 | 22.0 | 22.5 | 0.108 |
| | Sanitization | 11.9 | 13.3 | 0.108 | 24.3 | 20.4 | 0.125 | 23.7 | 22.9 | 0.134 | 25.2 | 24.9 | 0.142 |
| | **Ours** | **5.63** | **4.97** | **0.014** | **4.43** | **3.18** | **0.009** | **4.13** | **3.29** | **0.010** | **3.53** | **4.12** | **0.011** |
| LLaMA3-70B | W/ Attack | 96.5 | 94.2 | 0.924 | 91.0 | 88.0 | 0.856 | 96.8 | 97.6 | 0.958 | 98.6 | 95.6 | 0.923 |
| | FGSM | 19.4 | 20.3 | 0.117 | 32.3 | 27.3 | 0.215 | 37.3 | 35.6 | 0.234 | 35.6 | 32.7 | 0.194 |
| | FreeLB | 27.0 | 25.9 | 0.205 | 38.3 | 37.6 | 0.286 | 47.3 | 44.8 | 0.362 | 48.1 | 46.6 | 0.315 |
| | DPforward | 9.45 | 10.2 | 0.082 | 20.5 | 18.3 | 0.143 | 23.2 | 22.7 | 0.102 | 22.5 | 21.7 | 0.101 |
| | Sanitization | 11.9 | 13.7 | 0.103 | 19.3 | 17.7 | 0.085 | 24.0 | 23.7 | 0.120 | 27.0 | 26.4 | 0.108 |
| | **Ours** | **4.67** | **5.06** | **0.007** | **4.01** | **4.23** | **0.007** | **4.07** | **3.75** | **0.010** | **4.89** | **4.97** | **0.005** |
| Gemma2-9B | W/ Attack | 95.8 | 92.5 | 0.869 | 84.7 | 81.2 | 0.830 | 90.7 | 86.8 | 0.870 | 97.2 | 96.9 | 0.950 |
| | FGSM | 15.6 | 17.1 | 0.101 | 33.4 | 31.3 | 0.219 | 36.8 | 34.7 | 0.224 | 34.7 | 33.9 | 0.222 |
| | FreeLB | 33.8 | 28.8 | 0.221 | 41.8 | 40.9 | 0.265 | 53.7 | 50.1 | 0.372 | 49.9 | 48.2 | 0.314 |
| | DPforward | 9.54 | 10.7 | 0.072 | 20.0 | 19.8 | 0.128 | 20.7 | 19.5 | 0.113 | 19.9 | 17.6 | 0.099 |
| | Sanitization | 9.62 | 10.9 | 0.067 | 20.9 | 17.9 | 0.095 | 24.1 | 23.8 | 0.109 | 22.1 | 20.5 | 0.094 |
| | **Ours** | **4.20** | **4.50** | **0.007** | **4.38** | **4.42** | **0.012** | **4.81** | **4.63** | **0.009** | **4.50** | **4.60** | **0.012** |

Table 1: The overall performance of `Eguard` and other defense against embedding inversion attacks. Model: the type of embedding models, SST2, NLI, QR, and TS are the corresponding downstream datasets. Extended results are reported in the Appendix F.

**Evaluation on harmlessness.** To determine whether privacy protection compromises task performance, we evaluate the defended embeddings on sentiment analysis, natural language inference, question retrieval and text summarization. The results 7 reveal that while differential privacy methods reduce downstream accuracy due to heavy noise injection, and adversarial training slightly improves robustness but still lags behind, our approach retains over 98% of the original accuracy and ROUGE scores. These results indicate that our method successfully balances security and utility by decoupling sensitive information from the semantic representations without impairing discriminative power.

**Defense overhead.** We measure the training and inference overhead introduced by our projection network on T5 and MPNet using two NVIDIA RTX A6000 GPUs. On average, our method incurs a one-time training overhead of 1.6x–2.4x for MPNet and 2.1x–3.4x for T5 compared to undefended models (e.g., 16.3ms vs. 9.6ms per batch for SST2). Despite this additional cost, the overhead is acceptable given the substantial improvement in security and privacy.

## Evaluation on Robustness

**Robustness to embedding perturbations.** To evaluate robustness against typical perturbations in cloud storage or transmission, we add white noise, Gaussian noise, truncation, PCA and quantization to T5 embeddings on the SST2 dataset. As shown in Figure 3, noise slightly reduces downstream performance due to semantic loss, while PCA and truncation preserve stable results thanks to the redundancy of high-dimensional features. Under quantization, our defense still prevents more than 95 percent of token inversion and maintains over 89% downstream accuracy, indicating that such perturbations do not significantly weaken the protected space.

**Generalization to defenses unseen during training.** We examine generalization by training the projection network on one dataset or embedding model and evaluating on unseen ones. As shown in Figure 4, the left side of each subfigure presents results under the unseen dataset setting (e.g., training on NLI, testing on SST2 and QR), while the right shows the unseen embedding model setting (e.g., training on T5, testing on MPNet). The result on unseen datasets shows that using NLI for training yields better cross-dataset robustness due to its larger data scale. When transferring across embedding models such as training on T5 and testing on MPNet, both defense performance and task accuracy drop because of distinct feature spaces, yet our method still surpasses differential privacy baselines in maintaining both security and utility.

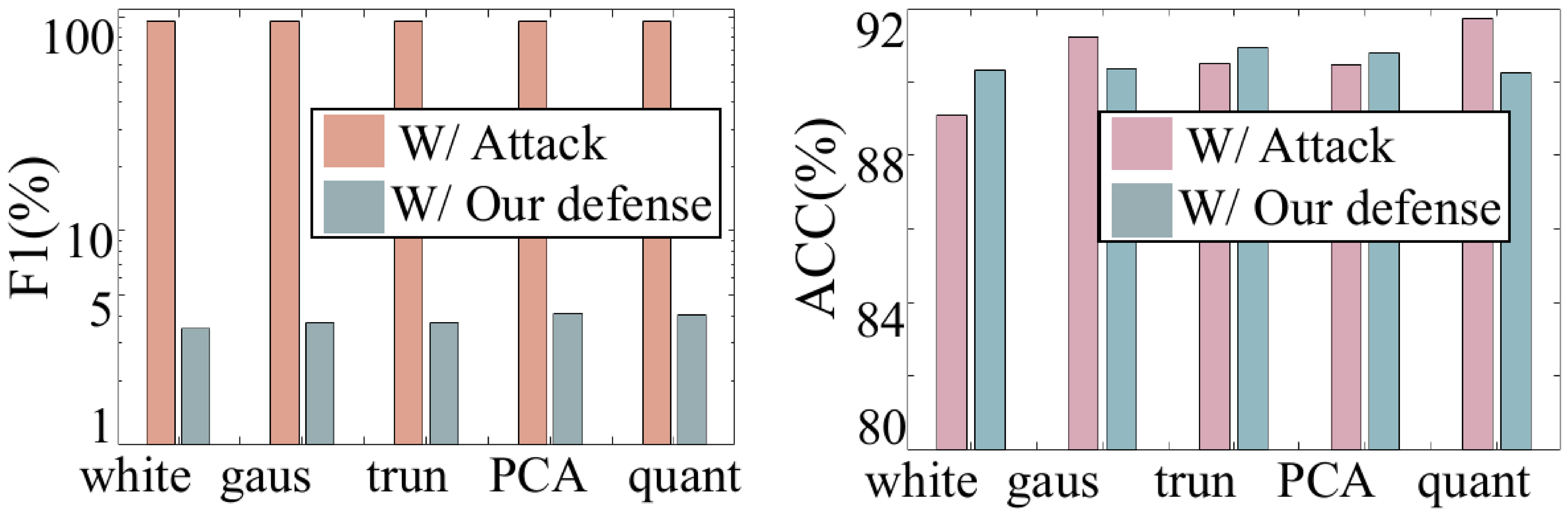

(a) Defense results. (b) Downstream performance.

Figure 3: The defense performance and downstream task performance under embedding perturbations.

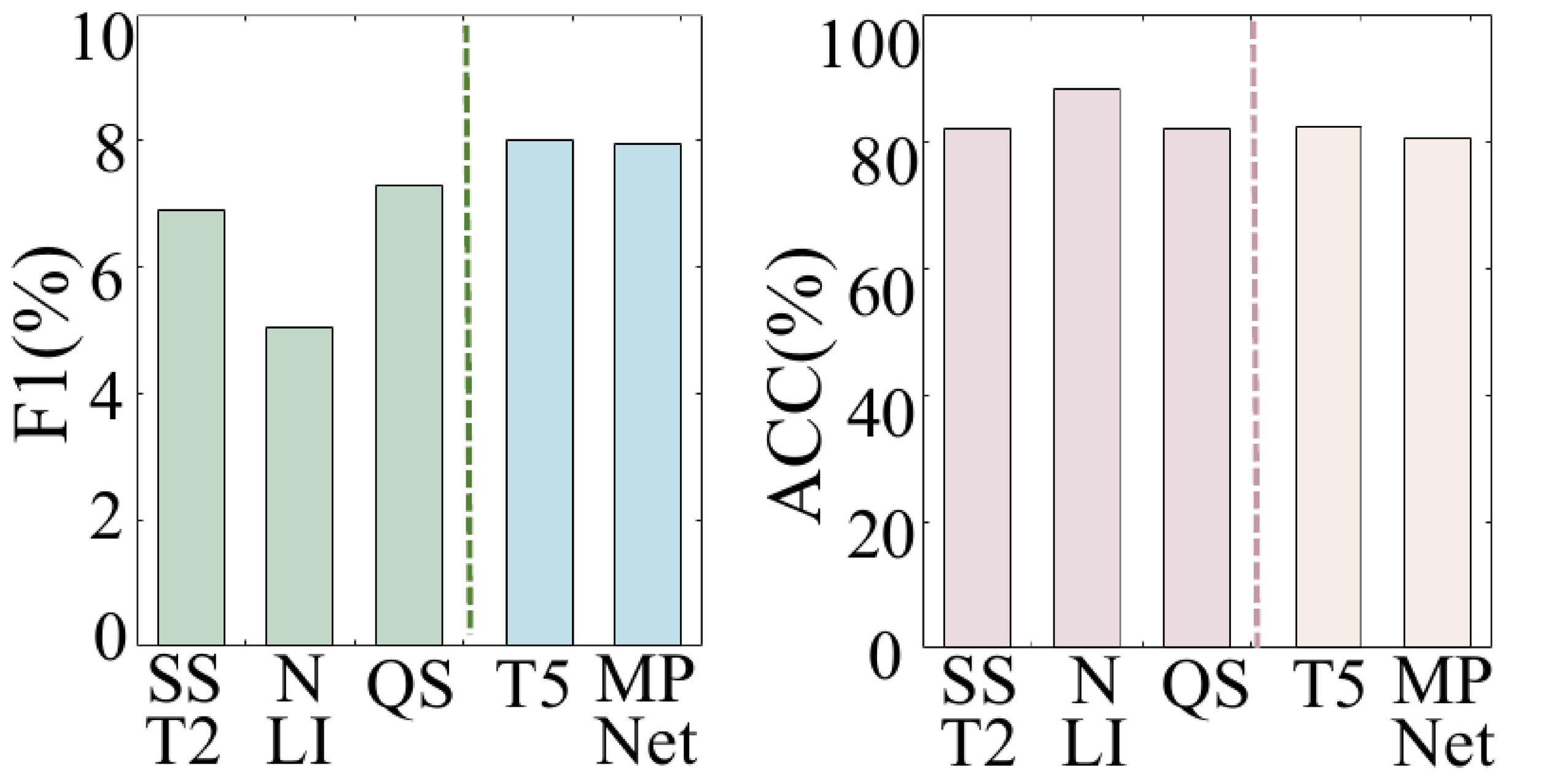

(a) Defense performance. (b) Task performance.

Figure 4: Robustness under unseen conditions.

| Model | Method | SST2(%) | NLI(%) | QR(%) | TS(%) |
|---|---|---|---|---|---|
| T5 | W/O Attack | **94.3** | **81.4** | **96.9** | **39.6** |
| | FGSM | 82.7 | 77.1 | 80.3 | 25.8 |
| | FreeLB | 84.4 | 80.1 | 84.4 | 28.6 |
| | DPforward | 60.7 | 46.9 | 48.9 | 19.4 |
| | Sanitization | 54.6 | 46.7 | 50.2 | 18.9 |
| | **Ours** | **93.8** | **81.8** | **96.7** | **38.3** |
| LLaMA2 | W/O Attack | **97.1** | **82.6** | **98.9** | **39.6** |
| | FGSM | 78.7 | 77.4 | 80.2 | 25.4 |
| | FreeLB | 82.1 | 72.3 | 83.1 | 20.3 |
| | DPforward | 60.8 | 56.9 | 60.1 | 18.9 |
| | Sanitization | 51.9 | 40.6 | 50.1 | 18.3 |
| | **Ours** | **96.8** | **81.8** | **98.1** | **38.7** |
| LLaMA3-70B | W/O Attack | **95.2** | **82.7** | **99.8** | **40.1** |
| | FGSM | 84.5 | 77.0 | 72.1 | 24.2 |
| | FreeLB | 81.8 | 77.5 | 74.1 | 22.3 |
| | DPforward | 71.1 | 69.6 | 63.1 | 20.1 |
| | Sanitization | 69.3 | 72.8 | 58.2 | 18.9 |
| | **Ours** | **94.8** | **80.3** | **97.9** | **39.8** |
| Gemma2-9B | W/O Attack | **93.4** | **80.8** | **96.3** | **38.9** |
| | FGSM | 85.8 | 73.4 | 82.7 | 23.1 |
| | FreeLB | 79.4 | 65.7 | 73.5 | 20.4 |
| | DPforward | 74.7 | 55.2 | 67.7 | 17.9 |
| | Sanitization | 63.5 | 51.4 | 63.7 | 17.7 |
| | **Ours** | **93.4** | **80.8** | **95.8** | **38.6** |

Table 2: Harmlessness evaluation of original and defended embeddings. ROUGE for summarization, accuracy for others. More results in Appendix G

**Attacking decoder.** We further replace GPT-2 with LLaMA2-7B, LLaMA3-8B and Gemma2-9B as inversion decoders targeting SST2 embeddings from T5, LLaMA2-7B and Gemma2-9B. Table 4 shows that all decoders display similar attack performance, but our defense consistently mitigates inversion attempts, demonstrating resilience across different large language model architectures.

## OpenAI embeddings

We evaluate embedding inversion attacks and defense strategies on OpenAI embeddings, which are also widely used for clustering, retrieval, and RAG applications. We use the OpenAI API with texts from the SST2 dataset to query three public models: text-embedding-3-small, text-embedding-3-large, and text-embedding-ada-002, with embedding dimensions of 1536, 1536, and 3072, respectively. Table 5 summarizes the performance of our defense compared to other baselines. Undefended embeddings show high F1 and Recall, confirming their vulnerability to inversion attacks. FGSM and FreeLB achieve limited protection, while DP-Forward and Sanitization offer moderate improvements. In contrast, our method achieves the lowest attack success rates, with F1 reduced to 5.28% for ada-002, 5.12% for 3-small, and 3.88% for 3-large, and Recall values ranging from 3.96% to 4.73%. These results demonstrate that our defense provides robust protection while maintaining strong embedding utility, outperforming all other approaches.

| Model | MPNet | | | T5 | | |
|---|---|---|---|---|---|---|
| Dataset | SST2 | NLI | TS | SST2 | NLI | TS |
| W/O Defense | 9.6ms | 10.4ms | 14.8ms | 6.3ms | 12.9ms | 14.9ms |
| W/ Defense | 16.3ms | 25.1ms | 24.7ms | 21.1ms | 34.2ms | 31.6ms |
| $\Delta$ | 6.7ms | 14.7ms | 9.9ms | 14.8ms | 21.3ms | 16.7ms |

Table 3: Overhead training comparison of `Eguard` per training batch versus the original undefended training in the same setting.

| Model | Defense | Attacking Decoder (F1%) | | |
|---|---|---|---|---|
| | | LLaMA2 | LLaMA3 | Gemma2 |
| T5 | W/O | 97.2 | 98.4 | 95.8 |
| | W/ | 3.70 | 3.91 | 4.03 |
| LLaMA2 | W/O | 92.9 | 94.4 | 91.8 |
| | W/ | 4.10 | 4.04 | 3.91 |
| Gemma2 | W/O | 93.5 | 98.2 | 98.6 |
| | W/ | 4.20 | 4.01 | 4.04 |

Table 4: The performance under different attacking decoders

## Ablation Studies

To evaluate the role of each component in our defense framework, we conduct ablation experiments by removing or modifying the mutual information loss and the projection

| Method | ada-002(%) | | 3-small(%) | | 3-large(%) | |
|---|---|---|---|---|---|---|
| | F1 | Recall | F1 | Recall | F1 | Recall |
| W/ Attack | 93.91 | 93.13 | 93.83 | 93.07 | 83.92 | 82.17 |
| FGSM | 13.28 | 11.37 | 13.54 | 11.37 | 13.41 | 10.26 |
| FreeLB | 14.19 | 12.11 | 13.95 | 11.06 | 14.57 | 10.19 |
| DPforward | 14.20 | 9.97 | 14.57 | 18.89 | 18.17 | 18.81 |
| Sanitization | 13.18 | 9.62 | 12.07 | 12.94 | 12.96 | 10.46 |
| Ours | 5.28 | 4.72 | 5.12 | 4.73 | 3.88 | 3.96 |

Table 5: Defense performance against embedding inversion attacks on OpenAI embeddings

network. The results are summarized in Figure 5.

**Impact of loss function.** We compare mutual information loss with mean squared error (MSE), cosine similarity, Mahalanobis distance, and adversarial loss. MSE and Mahalanobis losses reduce downstream accuracy to below 86%, and adversarial loss reaches about 92% but provides limited defense. Cosine similarity performs slightly better yet remains inferior to mutual information loss. The latter effectively reduces statistical dependency between original and protected embeddings, resulting in both stronger resistance to inversion and higher task performance. This confirms that capturing and minimizing information flow is more effective than merely optimizing distance-based metrics.

**Impact of projection network.** We further examine the influence of projection network architecture by comparing our 24-layer RoBERTa projection with LLaMA3, Transformer-XL, DeBERTa, XLNet, and a shallow MLP on SST2. While MLP fails to provide robust protection or maintain accuracy due to limited representational capacity, transformer-based architectures achieve stronger defense and comparable downstream performance. Deeper models like LLaMA3 and DeBERTa exhibit better generalization, suggesting that network depth and attention mechanisms are crucial for capturing complex semantic relations while introducing sufficient transformation to obscure sensitive content. Overall, our projection network strikes a balance between privacy protection and utility that simpler models cannot match.

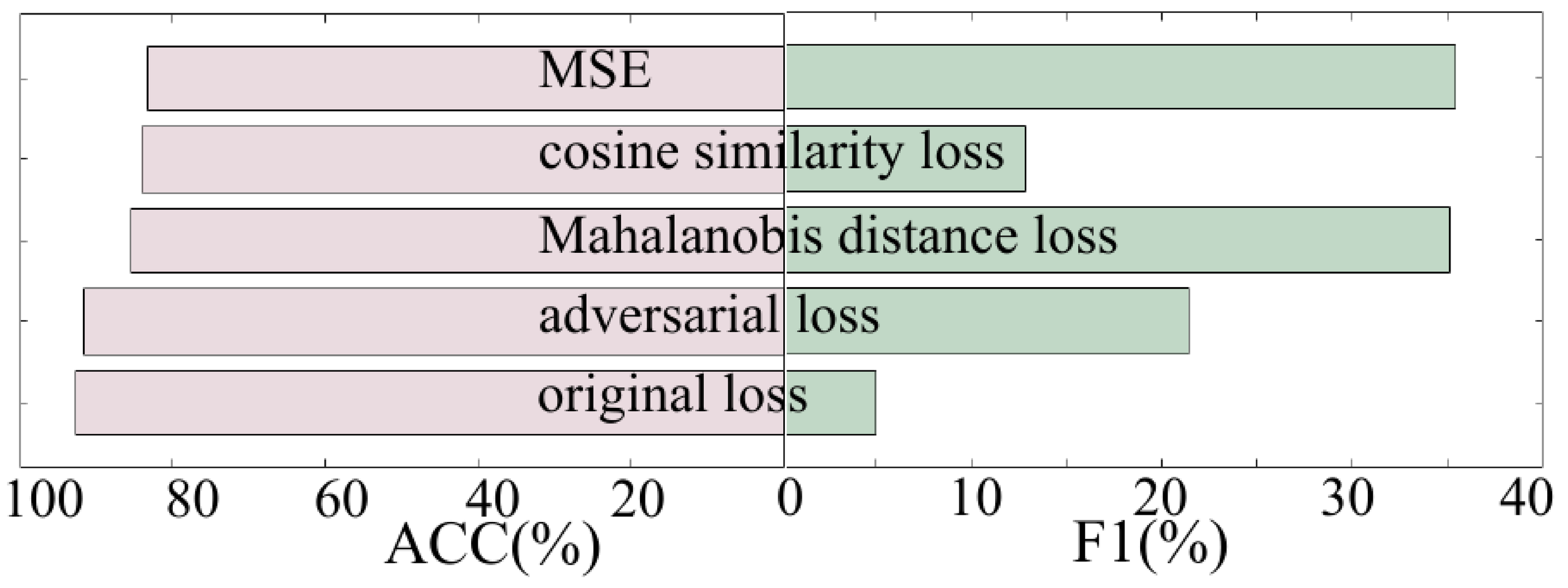

(a) Impact of loss function.

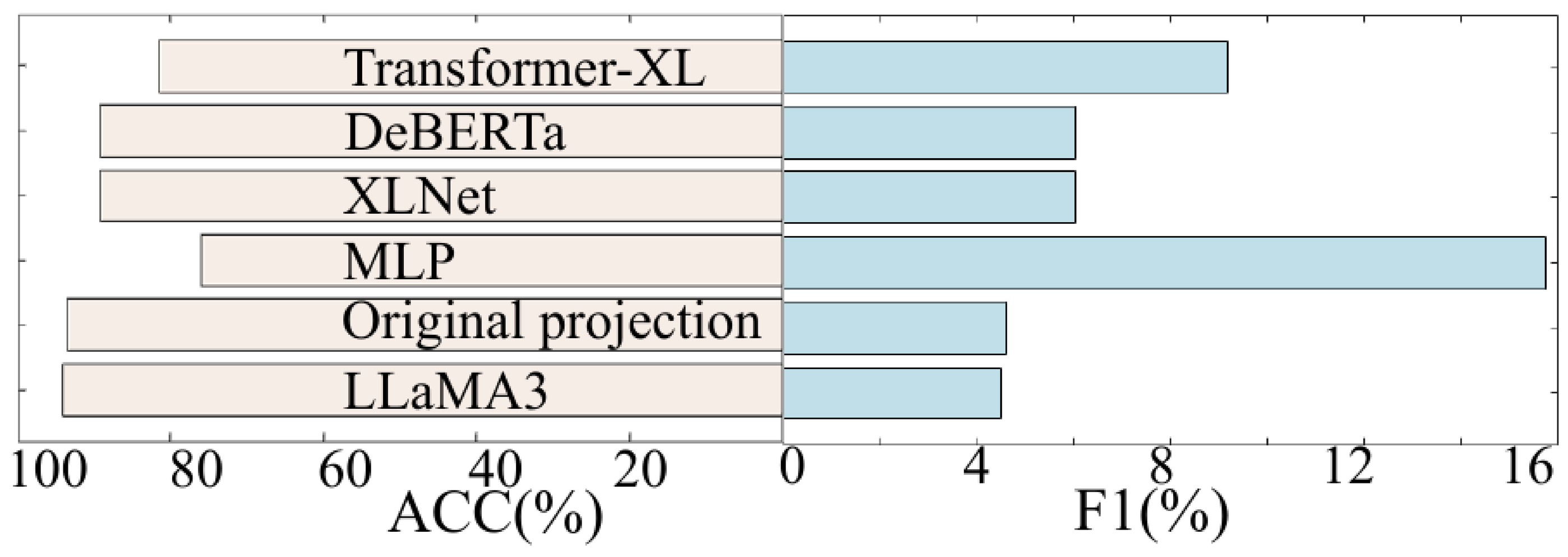

(b) Impact of projection network.

Figure 5: Ablation study results: (1) adjusting the loss function, and (2) investigating the impact of projection network architecture.

## Related Work

**Text embeddings are universal:** Text embeddings represent words, phrases, or documents as low-dimensional vectors, capturing semantic information for diverse NLP tasks (Ashkboos et al. 2024; Wang et al. 2023; Feng et al. 2020). Models such as Sentence-BERT (Reimers and Gurevych 2019), SimCSE (Gao, Yao, and Chen 2021), and Sentence-T5 (Ni et al. 2021a) fine-tune pre-trained encoders to generate embeddings for classification, QA, retrieval, and bitext mining. Recent efforts also extend embeddings across multiple domains, such as C-Pack for Chinese embeddings (Xiao et al. 2023), OpenAI embeddings for text and code (Neelakantan et al. 2022), and BGE for multilingual retrieval (Luo et al. 2024). Benchmark suites (MTEB (Muennighoff et al. 2022), SentEval (Conneau and Kiela 2018), BEIR (Thakur et al. 2021)) help compare embedding models on cross-domain tasks. Embeddings also enable anonymized storage of semantic data, facilitating model fine-tuning and personalized applications, as adopted by JINA AI (AI 2024), SingleStore (SingleStore 2023), and LangChain.

**Text embeddings security:** Embedding vectors are vulnerable to adversarial, membership inference, and reconstruction attacks (Song and Raghunathan 2020; Abdalla et al. 2020; Morris et al. 2023; Li, Xu, and Song 2023; Gu et al. 2023), which can expose sensitive information. To mitigate these risks, two main strategies exist: i) Differential privacy: InferDPT and DP-zero (Zhang et al. 2023; Tong et al. 2025) injects noise into gradients during optimization. Variants like d-privacy (Feyisetan et al. 2020) or DP-Forward (Du et al. 2023) perturb embeddings directly with controlled Gaussian noise. ii) Adversarial training: Models are optimized to resist adversarial perturbations (Liu et al. 2023; Wang et al. 2021), often improving the privacy-accuracy trade-off. Dai et al. introduce interpretable perturbations in the embedding space (Dai et al. 2019), while Yang et al. employ fast triplet metric learning to generate robust embeddings (Yang, Wang, and He 2022).

## Conclusion

In this paper, we propose `Eguard`, a novel defense method to achieve superior privacy-utility balance through dual-level mutual information optimization, addressing the critical issue of embedding inversion attacks. `Eguard` simultaneously minimizing global mutual information to prevent sensitive information leakage, while preserving task-relevant semantics through local contrastive learning. Extensive experiments across seven embedding models and four downstream tasks demonstrate that `Eguard` consistently achieves strong resistance to inversion with minimal impact on performance.

# Appendix

## A. Poof of Lemma 1

The embedding transformation process follows the Markov chain: $x \to e \to \hat{e} \to \hat{x}$, By applying the data processing inequality (DPI) to the Markov chain, we obtain (Zou et al. 2024):

$$I(x, \hat{x}) \leq I(x, \hat{e}) \leq I(e, \hat{e}). \tag{11}$$

By optimizing the projection network to minimize $I(x, \hat{e})$, we indirectly constrain $I(x, \hat{x})$, thereby limiting the adversary's ability to reconstruct $x$. In an information-theoretic sense, the projection network introduces an information bottleneck, discarding sensitive details while retaining only the minimal task-relevant statistical knowledge. Consequently, as $I(x, \hat{e})$ approaches zero, the conditional entropy $H(x|\hat{e})$ increases, leading to maximal uncertainty for the attacker. This proves that controlling mutual information provides a defense against inversion attacks.

## B. Poof of Lemma 2

Let $S^* = \log \frac{d\mathbb{P}_{z\hat{e}}}{d(\mathbb{P}_z \otimes \mathbb{P}_{\hat{e}})}$, the optimal critic function. By definition, $S^*$ satisfies:

$$\mathbb{E}_{\mathbb{P}_{z\hat{e}}}[S^*] = I(z, \hat{e}), \quad \mathbb{E}_{\mathbb{P}_z \otimes \mathbb{P}_{\hat{e}}}[e^{S^*}] = 1. \tag{1}$$

The estimation gap $\Delta = |I(z, \hat{e}) - I_{\text{global}}(z, \hat{e})|$ is bounded by:

$$\Delta \leq \underbrace{\mathbb{E}_{\mathbb{P}_{z\hat{e}}}[S^* - S_\theta]}_{\text{Approximation Error}} + \underbrace{\mathbb{E}_{\mathbb{P}_z \otimes \mathbb{P}_{\hat{e}}}[e^{S_\theta} - e^{S^*}]}_{\text{Regularization Error}}. \tag{2}$$

We proceed in two cases (Belghazi et al. 2018):

**Case 1: Bounded $S^* \leq M$:** By the universal approximation theorem (Hornik, 1989), for $\epsilon/2 > 0$, there exists $S_\theta$ such that:

$$\mathbb{E}_{\mathbb{P}_{z\hat{e}}}[S^* - S_\theta] \leq \epsilon/2, \quad \mathbb{E}_{\mathbb{P}_z \otimes \mathbb{P}_{\hat{e}}}[S^* - S_\theta] \leq \epsilon e^{-M}/2. \tag{3}$$

Using the Lipschitz continuity of $e^x$ on $(-\infty, M]$:

$$\mathbb{E}_{\mathbb{P}_z \otimes \mathbb{P}_{\hat{e}}}[e^{S^*} - e^{S_\theta}] \leq e^M \mathbb{E}_{\mathbb{P}_z \otimes \mathbb{P}_{\hat{e}}}[S^* - S_\theta] \leq \epsilon/2. \tag{4}$$

Substituting (3) and (4) into (2):

$$\Delta \leq \epsilon/2 + \epsilon/2 = \epsilon. \tag{5}$$

**Case 2: Unbounded $S^*$:** For $\epsilon/4 > 0$, partition $\Omega$ into $\{S^* \leq M\}$ and $\{S^* > M\}$. By the Dominated Convergence Theorem, choose $M$ such that:

$$\mathbb{E}_{\mathbb{P}_{z\hat{e}}}[S^* \mathbb{1}_{\{S^* > M\}}] \leq \epsilon/4, \quad \mathbb{E}_{\mathbb{P}_z \otimes \mathbb{P}_{\hat{e}}}[e^{S^*} \mathbb{1}_{\{S^* > M\}}] \leq \epsilon/4. \tag{6}$$

On $\{S^* \leq M\}$, apply Case 1 to get $S_\theta \leq M$ satisfying:

$$\begin{aligned} &\mathbb{E}_{\mathbb{P}_{z\hat{e}}}[|S^* - S_\theta| \mathbb{1}_{\{S^* \leq M\}}] \leq \epsilon/2, \\ &\mathbb{E}_{\mathbb{P}_z \otimes \mathbb{P}_{\hat{e}}}[|S^* - S_\theta| \mathbb{1}_{\{S^* \leq M\}}] \leq \epsilon e^{-M}/4. \end{aligned} \tag{12}$$

Combining (6) and (7):

$$\mathbb{E}_{\mathbb{P}_z \otimes \mathbb{P}_{\hat{e}}}[e^{S^*} - e^{S_\theta}] \leq e^M \cdot \epsilon e^{-M}/4 + \epsilon/4 = \epsilon/2. \tag{8}$$

Thus, $\Delta \leq \epsilon/2 + \epsilon/2 = \epsilon$.

This proves that, regardless of whether $S^*$ is bounded or unbounded, the estimation error is **uniformly controlled** within $\epsilon$. As a result, the neural network estimator $I_{\text{global}}(z, \hat{e})$ **converges almost surely** to $I(z, \hat{e})$ as the sample size increases, thereby ensuring the consistency of the learned mutual information measure.

## C. Proof of Lemma 3

Define the gap $\Delta = I_{\text{local}} - I(k, a)$:

$$\begin{aligned} \Delta &= \mathbb{E}_{p(a)} \left[ \mathbb{E}_{p(k|a)}[\log p(k)] - \mathbb{E}_{p(k)}[\log p(k)] \right] \\ &= \int p(a) \left( \int p(k|a) \log p(k) \, dk - \int p(k) \log p(k) \, dk \right) da. \end{aligned} \tag{13}$$

By Jensen's inequality for the concave log-function, $\mathbb{E}_{p(k|a)}[\log p(k)] \leq \log \mathbb{E}_{p(k|a)}[p(k)] = \log p(k)$, ensuring $\Delta \geq 0$. Equality holds when $p(k|a) = p(k)$, i.e., independence (Cheng et al. 2020).

# D. Embeddings Models

To conduct embedding inversion attacks and validate our defense mechanisms, we have chosen four cutting-edge embedding models. The parameters of these models remain frozen, and we utilize the pre-trained weights as provided in their respective original GitHub repositories.

**T5** (Ni et al. 2021b), introduced by Google, is an innovative encoder-decoder model that has been pre-trained on a diverse and voluminous dataset, leveraging transfer learning for enhanced performance. In our work, we utilize the encoder of the T5-large model to project sentences into a 768-dimensional vector space. The T5-large model has a total of 770 million parameters.

**RoBERTa** (Liu et al. 2019), an optimized variant of the BERT model, undergoes pretraining with a focus on the masked language modeling objective. The pretrained RoBERTa-large model encodes the input sentence into a 1024 dimension vector and is fine-tuned in on a 1B sentence pairs dataset. The RoBERTa-large model parameters are 1360 MB in size.

**MPNet** (Song et al. 2020), introduced by Microsoft, combines the strengths of masked language modeling from BERT and permuted language modeling from XLNet, resulting in improved performance across a range of downstream benchmark tasks. We use the encoder of MPNet to produce 768-dimensional embeddings, which are 420 MB in size.

**LLaMA** (Touvron et al. 2023), proposed by Meta AI, stands as a state-of-the-art foundational large language model that is tuned by using supervised fine-tuning and reinforcement learning with human feedback. We leverage the LLaMA2-7B, LLaMA3-8B, and LLaMA3-70B as our encoder to generate 4096 dimensional embeddings.

**Gemma** (Team 2024), proposed by Google, is designed for text generation tasks like question answering, summarization, and reasoning. These models are small and can be used on devices with limited resources, such as laptops and desktops. We use Gemma2-9B as the encoder to obtain the 3584-dimensional embedding vector.

# E. Datasets and Language Tasks

We introduce the text dataset that will be encoded by the embedding model and its corresponding language understanding tasks, including sentiment analysis, natural language inference, question retrieval, and text summarization. For brevity, we refer to each dataset by its respective acronym.

**SST: Sentiment Analysis.** The Stanford sentiment treebank (SST2) (Socher et al. 2013) is a corpus containing 67,000 sentences sourced from movie reviews, each meticulously annotated with human-derived sentiment labels. The goal of sentiment analysis is to determine the emotional tone of a sentence, categorizing it as positive or negative, which is a type of classification problem in language processing.

**NLI: Natural Language Inference.** The task of natural language inference (NLI) involves determining the logical relationship between two sentences—a premise and a hypothesis. Specifically, the goal is to classify the relationship as entailment (the premise supports the hypothesis), contradiction (the premise contradicts the hypothesis), or neutrality (there is no evident logical correlation). The Stanford natural language inference (SNLI) (Glockner, Shwartz, and Goldberg 2018) and multi-genre natural language inference (MultiNLI) (Kim, Kang, and Kwak 2019) corpus are prominent resources developed for advancing machine learning models' capabilities in sentence comprehension. In our experimental setup, we consolidate the SNLI and MultiNLI datasets into a unified NLI dataset, which encompasses premise sentences from a diverse range of sources, including transcriptions, reports, speeches, letters, and works of fiction. This amalgamated dataset comprises 1 million sentence pairs.

**QR: Question Retrieval.** The task of question retrieval involves identifying the most relevant question within a vast corpus containing thousands of questions. As an information retrieval task, its objective is to match a given question with its closest duplicate in the corpus. The dataset used for this purpose is the Quora duplicates questions dataset (Wang, Hamza, and Florian 2017), which comprises 500,000 questions covering a wide array of topics and issues originally posted on Quora.

**TS: Text Summarization.** Text Summarization (TS) entails the compression of extensive textual information into succinct summaries. This dataset amalgamates content from USAToday obtained from CNN (Zhong et al. 2019) and SoLSCSum sourced from Yahoo News (Nguyen et al. 2016). It encompasses roughly 10k sentences extracted from a corpus of over 200 articles, each paired with a gold-standard summary crafted by human annotators. Analyzing the TS dataset reveals that the source articles exhibit an average word count of 500, comprising roughly 22 sentences, while the summaries are significantly abbreviated, averaging a length of 20 words.

# F. Parameter Settings

To ensure reproducibility and clarity, we provide the detailed parameter configurations used in our experiments, covering both the proposed Eguard defense mechanism and the baseline comparative methods.

**Eguard parameters.** The projection network in Eguard is implemented as a 24-layer RoBERTa encoder, serving as an intermediate transformation module between raw embeddings and the secured latent space. The model is optimized using the Adam optimizer with a learning rate of $3e-2$ and a batch size of 8. The trade-off coefficients for mutual information regularization are set to $\alpha = 1.0$ for the global mutual information term and $\beta = 0.1$for the local mutual information term, as determined through cross-validation on the development set.

**Differential privacy baselines.** For DP-based defenses (DPforward and Sanitization), we adopt the privacy budget $\epsilon \in \{0.1, 1, 10\}$ and clipping norm $C \in \{0.5, 1.0, 2.0\}$. The reported results correspond to the optimal configuration, i.e., $\epsilon = 10.0, C = 1.0$ which balances defense strength and downstream task performance.

**Adversarial training baselines.** For adversarial training methods (FGSM and FreeLB), we apply perturbations to the input embedding layer. FreeLB is run with 3 ascent steps and a perturbation radius of 0.5, while FGSM uses a single-step perturbation with a step size of 0.25. These parameters are chosen following prior work to provide robust baselines without compromising model stability.

## G. Rest results of Table 1

| Model | Method | SST2 | | | NLI | | | QR | | | TS | | |
|---|---|---|---|---|---|---|---|---|---|---|---|---|---|
| | | F1(%) | Recall(%) | BLEU | F1(%) | Recall(%) | BLEU | F1(%) | Recall(%) | BLEU | F1(%) | Recall(%) | BLEU |
| RoBERTa | W/ Attack | 93.9 | 93.2 | 0.836 | 82.4 | 82.7 | 0.831 | 98.2 | 97.9 | 0.981 | 95.6 | 94.8 | 0.912 |
| | FGSM | 17.7 | 17.8 | 0.112 | 36.7 | 30.1 | 0.278 | 36.8 | 34.8 | 0.212 | 37.5 | 36.1 | 0.235 |
| | FreeLB | 18.5 | 17.6 | 0.104 | 35.7 | 30.2 | 0.278 | 51.3 | 48.4 | 0.238 | 49.8 | 48.7 | 0.320 |
| | DPforward | 14.3 | 10.1 | 0.014 | 23.7 | 18.9 | 0.257 | 24.4 | 20.5 | 0.147 | 22.1 | 20.1 | 0.131 |
| | Sanitization | 14.6 | 13.6 | 0.083 | 24.7 | 19.8 | 0.125 | 24.0 | 22.3 | 0.137 | 25.9 | 24.7 | 0.109 |
| | **Ours** | **4.45** | **4.41** | **0.019** | **3.15** | **4.21** | **0.014** | **2.98** | **3.23** | **0.013** | **3.21** | **4.12** | **0.008** |
| MPNet | W/ Attack | 93.9 | 93.4 | 0.837 | 83.2 | 83.3 | 0.822 | 98.8 | 97.9 | 0.980 | 96.1 | 95.2 | 0.906 |
| | FGSM | 17.4 | 17.6 | 0.115 | 36.4 | 29.8 | 0.269 | 37.0 | 34.8 | 0.221 | 37.8 | 36.8 | 0.212 |
| | FreeLB | 22.7 | 21.5 | 0.145 | 29.3 | 24.4 | 0.178 | 50.8 | 49.0 | 0.304 | 49.0 | 48.7 | 0.328 |
| | DPforward | 13.7 | 10.3 | 0.015 | 23.1 | 18.8 | 0.138 | 26.5 | 24.1 | 0.167 | 22.8 | 20.8 | 0.167 |
| | Sanitization | 9.61 | 17.1 | 0.036 | 17.8 | 15.0 | 0.088 | 23.8 | 21.8 | 0.118 | 24.6 | 23.4 | 0.116 |
| | **Ours** | **5.15** | **4.43** | **0.012** | **4.55** | **4.13** | **0.011** | **4.12** | **4.21** | **0.009** | **4.31** | **5.24** | **0.010** |
| LLaMA3-8B | W/ Attack | 94.2 | 93.6 | 0.839 | 91.4 | 88.5 | 0.817 | 89.2 | 88.5 | 0.870 | 97.2 | 94.3 | 0.937 |
| | FGSM | 13.3 | 15.8 | 0.097 | 24.9 | 21.4 | 0.149 | 36.8 | 34.7 | 0.227 | 40.3 | 38.3 | 0.272 |
| | FreeLB | 25.6 | 21.7 | 0.136 | 44.6 | 44.3 | 0.352 | 45.9 | 40.8 | 0.303 | 49.8 | 47.2 | 0.327 |
| | DPforward | 8.75 | 11.5 | 0.046 | 24.0 | 22.7 | 0.172 | 21.5 | 18.9 | 0.121 | 21.9 | 20.8 | 0.105 |
| | Sanitization | 9.57 | 10.7 | 0.066 | 25.8 | 24.3 | 0.113 | 23.2 | 23.8 | 0.139 | 25.0 | 24.2 | 0.109 |
| | **Ours** | **3.26** | **2.88** | **0.003** | **4.27** | **4.56** | **0.013** | **3.73** | **4.38** | **0.006** | **4.38** | **4.48** | **0.008** |

Table 6: The overall performance of `Eguard` and other defense against embedding inversion attacks. Model: the type of embedding models, W/ Attack: the embedding inversion attacks on embeddings. SST2, NLI, QR, and TS are the corresponding downstream datasets.

## H. Rest results of Table 2

| Model | Method | SST2(%) | NLI(%) | QR(%) | TS(%) |
|---|---|---|---|---|---|
| RoBERTa | W/O Attack | **93.8** | **81.8** | **97.9** | **38.9** |
| | FGSM | 70.2 | 79.4 | 80.1 | 21.3 |
| | FreeLB | 77.1 | 70.5 | 80.8 | 19.6 |
| | DPforward | 59.4 | 48.2 | 58.5 | 16.8 |
| | Sanitization | 47.3 | 43.2 | 60.3 | 17.7 |
| | **Ours** | **94.0** | **80.8** | **96.8** | **37.6** |
| MPNet | W/O Attack | **93.8** | **79.7** | **99.1** | **37.8** |
| | FGSM | 43.1 | 79.2 | 79.7 | 21.5 |
| | FreeLB | 78.7 | 78.9 | 51.6 | 20.7 |
| | DPforward | 62.1 | 52.1 | 83.3 | 18.9 |
| | Sanitization | 53.2 | 44.8 | 51.6 | 17.8 |
| | **Ours** | **93.8** | **79.8** | **96.8** | **37.6** |
| LLaMA3-8B | W/O Attack | **94.9** | **85.0** | **97.9** | **39.8** |
| | FGSM | 81.3 | 77.9 | 83.8 | 20.1 |
| | FreeLB | 88.6 | 74.6 | 81.5 | 19.8 |
| | DPforward | 70.1 | 64.7 | 72.9 | 16.7 |
| | Sanitization | 62.4 | 58.8 | 69.7 | 16.3 |
| | **Ours** | **93.2** | **81.1** | **97.9** | **39.6** |

Table 7: Evaluation of Harmlessness: Comparison of downstream task performance between original embeddings and embeddings subjected to different defense mechanisms. W/O attack denotes the original embedding without the influence of any attack or defense. Note that ROUGE is used for text summarization, and accuracy for the other tasks.